# X-ray photoelectron spectroscopy, Magnetotransport and Magnetization study of $Nb_2PdS_5$ superconductor

Reena Goyal[1,2], A. K. Srivastava[1], Monu Mishra[1,2], Govind Gupta[1], Rajveer Jha[3] and V. P. S. Awana[1,*]

[1]*CSIR- National Physical Laboratory, Dr. K. S. Krishnan Marg, New Delhi-110012, India*
[2]*AcSIR - Academy of Scientific & Innovative Research - National Physical Laboratory (CSIR) New Delhi-110012, India*
[3]*Department of Physics, Tokyo Metropolitan Univ. Hachioji-shi, Minami Osawa, 1-1, Tokyo 192-0397, Japan*

**Abstract**

In the present report, we investigate various properties of the $Nb_2PdS_5$ superconductor. Scanning electron microscopy displayed slabs like laminar growth of $Nb_2PdS_5$ while X-ray photoelectron spectroscopy exhibited the hybridization of Sulphur (2p) with both Palladium (3d) and Niobium (3d). High field (140kOe) magneto-transport measurements revealed that superconductivity ($T_c^{onset}$ =7K and $T_c^{\rho=0}$ = 6.2K) of the studied $Nb_2PdS_5$ material is quite robust against magnetic field with the upper critical field ($H_{c2}$) outside the Pauli paramagnetic limit. Thermally activated flux flow (TAFF) of the compound showed that resistivity curves follow Arrhenius behaviour. The activation energy for $Nb_2PdS_5$ is found to decrease from 15.15meV at 10kOe to 2.35meV at 140kOe. Seemingly, the single vortex pinning is dominant in low field regions, while collective pinning is dominant in high field region. The temperature dependence of AC susceptibility confirmed the $T_c$ at 6K, further varying amplitude and frequency showed well coupled granular nature of superconductivity. The lower critical field ($H_{c1}$) is extracted from DC magnetisation measurements at various $T$ below $T_c$. It is found that $H_{c1}(T)$ of $Nb_2PdS_5$ material seemingly follows the multiband nature of superconductivity.

*Corresponding Author
Dr. V. P. S. Awana, Principal Scientist,
E-mail: awana@mail.npindia.org
Ph. +91-11-45609357, Fax-+91-11-45609310,
Homepage: awanavps.wenbs.com

**Introduction**

Recently, (Nb/Ta)$_2$Pd(S/Se/Te)$_5$ series is gathering the interest of researchers due to their intriguing properties like robustness against the magnetic field and formation of the linear Pd(S/Se/Te) chain along one direction [1-11]. Structurally, all the compounds of series (Nb/Ta)$_2$Pd(S/Se/Te)$_5$ crystallize in monoclinic structure with space group C2/m [1, 8-11]. In 2013, superconductivity was reported by Q. Zhang et al. in Nb$_2$PdS$_5$ with transition temperature $T_c$ =7K and upper critical field of about 370kOe [1]. Due to the high upper critical field and low dimensional structure, these compounds attracted the attention of the scientific community in a big way [1-4]. The variation in the upper critical field, with Pd site substitution of Ni, and Pt has been studied to highlight the possible role of spin orbitcoupling in this compound.

The band structure calculations revealed that Fermi surface of Nb$_2$PdS$_5$ consists of a set of electron-like flat sheet and hole like cylinders [1]. The multiband nature of superconductivity in Nb$_2$PdS$_5$ has already been confirmed by the results of upper critical field, specific heat and angle-resolved photoelectron spectroscopy [1, 13, 14]. Detailed electronic structure calculations on these compounds also do suggest the multiband nature of these compounds [15]. Worth mentioning is the fact that the low dimensional conduction chains, robustness of superconductivity and the multiband nature of electronic states are intrinsic to whole series of (Nb/Ta)$_2$Pd(S/Se/Te)$_5$ compounds [1-15].

In this shortarticle, we present some yet missing but interesting properties of one of the (Nb/Ta)$_2$Pd(S/Se/Te)$_5$ series compound, i.e., Nb$_2$PdS$_5$. Here, we present thermally activated flux flow energy, thetemperature dependence of AC susceptibility with varying amplitude and frequency along with X-ray photoelectron spectroscopy (XPS) of Nb$_2$PdS$_5$ sample. The temperature dependence of lower critical field is suggestive of the multiband nature of superconductivity in studied Nb$_2$PdS$_5$. XPS studies exhibited the hybridization of Sulphur (2p) with both Palladium (3d) and Niobium (3d). High field magneto transport studies revealed that single vortex pinning is dominant in the low field region, while the collective pinning is dominant in high field regions.

**Experimental Details**

The Nb$_2$PdS$_5$ sample was synthesized using standard solid state reaction route. The details of synthesis and structural characterization of Nb$_2$PdS$_5$ are given elsewhere [13]. Scanning electron microscopy (SEM) was carried using ZEISS-EVO scanning electron microscope. High resolution transmission electron microscopy was employed using HRTEM, modeloperated at300 kV electron accelerating voltage.X-ray photoelectron was carried using Omicron multiprobe surface analysis system equipped with Monochromatic Al-Kα source having excitation energy of 1486.7eV. The electrical and Magnetic measurements were recorded using a quantum design physical property measurement system (PPMS) equipped with thesuperconducting magnet of 140 kOe. The AC magnetization measurements were also performed on PPMS in zero DC field and AC amplitude of 5-15Oe along with varying frequency from 333 to 9999Hz.

**Results and Discussion**

The room temperature XRD of $Nb_2PdS_5$ revealed monoclinic structure with space group C2/m, details are reported by us elsewhere [13]. The SEM image of the bulk polycrystalline $Nb_2PdS_5$ compound shows layered growth with slab like grains structure [see Fig. 1]. The Energy Dispersive X- ray spectroscopy (EDX) of the $Nb_2PdS_5$ sampleis also shownalong side in Fig.1. The EDX analysis is done on small portion of SEM image of same sample. It is clear that there is no other element present in sampleexcept Nb, Pd and S, further the compound was found to be near stoichiometric i.e., composition close to $Nb_2PdS_5$.

From the contrast developed over a region of about 1.5μm and 1.2μm in dimension (Fig. 2a), it is clearly visible that there are several grain boundaries in the microstructure, appearing as dark thin lines presumably due to disorientations between them.Although the microscopic image as depicted in Fig. 2a reveals elongated grey level contrast evolved due to different interfaces of grain boundaries, the further understanding of the atomic scale arrangements were dealt in different regions of the microstructure. As an illustrative example, inset 'A' in Fig. 2a delineates the atomic scale arrangement of a set of crystallographic planes of a monoclinic crystal with interplanar separation of $d$ = 0.33nm with hkl indices of 110 (space group: C2/m, [2, 4]) from a smoother region, marked as 'A' over in Fig.2a. Further in a detailed investigation at high magnifications (Fig.2b), normally a layered structure has been observed with a periodic separation between them. This periodicity is quite consistent throughout in the microstructure having a self repeat of an interlayer separation of about 1.5nm (Fig. 2(b)), corresponding to set of planes with hkl:001 of a monoclinic crystal structure of $Nb_2PdS_5$ [2]. A few of such repeated layers are marked with a set of arrows in Fig. 2(b). A corresponding selected area electron diffraction pattern (SAEDP) recorded from the region in Fig.2(b), replicates the layered atomic arrangement of real space in sharp and discrete diffraction spots in reciprocal space, as depicted in Fig2(c). In addition, two diffused rings in the SAEDP (Fig. 2(c)), marked as 1 and 2, distinguishes further that the overall microstructure is layered with a columnar chain-like appearance. However due to presence of large networking of elongated grain boundaries (Fig. 2(a)), the material is intrinsically a polycrystalline. The HRTEM image and SAEDP (Fig.2) are in corroboration with the observations reported in literature [2, 4]. Based on a previous report [4] it is presumably the monoclinic (space group C2/m) of $Nb_2PdS_5$ basic unit forming a chained arrangement along the [010] direction or in other words b-axis of a monoclinic unit cell sharing their face and edges and subsequent periodic layered-stacking in three-dimensional crystal structure is evolved. It is further inferred from the present and previous reports [2, 4] that the lattice constants of such monoclinic crystal of $Nb_2PdS_5$ with a, b, and c are 1.2, 0.34, and 1.5nm, respectively.

The resultant XPS core level spectra of Niobium (Nb), palladium (Pd) and Sulphur is given in Fig3. The Nb(3d) XPS (Fig. 3(a)) spectra level splits into two intense peaks that correspond to $3d_{5/2}$ and $3d_{3/2}$ core levels at binding energy positions of 203.1 and 205.9eV, respectively. Due to spin orbit coupling, the splitting of 3d peak of Nb was obtained to be 2.8eV, which is in agreement with our previous report on $Nb_2PdSe_5$ [16]. Further, the $3d_{5/2}$ and $3d_{3/2}$ core level spectra were deconvoluted into four main peaks. The positions of these peaks are at 203.04, 203.9 and 205.9, 206.9eV for Nb $3d_{5/2}$ and $3d_{3/2}$ respectively. The peaks positioned at 203.04 and 205.9 eV are assigned to the hybridization of Nb with S [17]. The other peaks at 203.9 and 206.9 eV are assigned to the formation of niobium oxide [16]. Similarly, Fig. 3(b) shows the splitting of Pd 3d XPS core level spectra into two intense peaks ($3d_{5/2}$ and $3d_{3/2}$) at binding energy positions of 336.3 and 341.5eV, respectively. Here, the splitting of 3d peak of Pd due to spin orbit coupling comes out to be 5.2eV [16]. Further these two peaks were deconvoluted into four main peaks at 336.2, 337.3, 341.5 and 342.5eV, respectively. The peaks located at 336.2 and 341.5eV represent the hybridization of Pd with sulphur, while peaks at position 337.3 and 342.5eV suggest the oxidation of Pd [18,19]. Also, the 2p core level spectra of Sulphur (S) could be deconvoluted into four peaks at 160.3, 161.1, 161.8 and 162.6eV respectively, shown in Fig.3 (c). The two fitted peaks at 160.3eV and 161.8eV (S $2p_{3/2}$ and S $2p_{1/2}$) correspond to the hybridization of Niobium with Sulphur [20] and the other two peaks correspond to the positions at 161.1eV and 162.6eV (S $2p_{3/2}$ and S $2p_{1/2}$) represent the hybridization of Pd with Sulphur [21].

Figure 4 (a) represents temperature dependence of electrical resistivity of $Nb_2PdS_5$ in the temperature range of 2K to 300K. The $Nb_2PdS_5$ sample shows superconductivity with $T_c^{onset}$ = 7.0K and $T_c^{\rho=0}$ = 6.2K. Inset of Fig.4 (a) shows temperature dependence resistivity under different magnetic field. Clearly, the superconducting transition temperature $T_c$ shifts towards lower temperature with increasing magnetic field. Seemingly, the studied $Nb_2PdS_5$ is quite robust against magnetic field. The detailed analysis of resistivity with and without magnetic field for $Nb_2PdS_5$ sample is given elsewhere [3, 13].

Here, we mainly focus on thermally activated flux flow (TAFF) analysis of the magneto transport properties of the bulk polycrystalline $Nb_2PdS_5$ compound in its superconducting regime, i.e., below the onset of $T_c$. According to the TAFF theory, the relationship between $\ln\rho$ and $1/T$ can be described using Arrhenius relation, $\ln\rho(T, H) = \ln\rho_o(H)-U_o(H)/T$, where $\ln\rho_o(H) = \ln\rho_{of} + U_o(H)/T_c$ is temperature independent and $U_o$ is apparent activated energy [22]. Therefore, from the above mentioned equation, we can see that, in TAFF region, the variation of $\ln\rho(T, H)$ vs. $1/T$ should be linear. Fig.4 (b) represents well fitted experimental data points using Arrhenius relation. Clearly, the $\ln\rho(T, H)$ is varying linearly with temperature, implying that the behaviour of temperature dependence of thermally activated energy (TAE) is linear, i.e., $U(T,H)=U_o(H)(1-T/T_c)$ [23]. As shown in the same figure the fitted $\ln\rho(T, H)$ lines extrapolate to same temperature assigned as $T_{cross}$. The value of $T_{cross}$ in $Nb_2PdS_5$ sample comes out to be 8K which is nearly equal to $T_{c,}^{onset}$, obtained from $\rho$ - $T$ measurements. Fig.4 (c) represents plot of $\ln\rho_o$ vs. $Uo$, which shows $\ln\rho_o$ is varying linearly with $Uo$. As shown in Fig.4 (d), $U_o(H)$ curve follows power law and is given by $Uo(H) \propto H^{-\alpha}$. This curve shows that the activation energy for $Nb_2PdS_5$ sample decreases from 15.15meV at 10kOe to 2.35 meV at 140kOe. The curve was fitted separately in two distinct regions defined as (i) 1kOe = H≤40kOe & (ii) 80kOe= H ≤140kOe. The extracted value of $\alpha$ = 0.43 for H≤40kOe and $\alpha$ = 1.32 for H>40 kOe. In low field region, single vortex pinning dominates since the variation of $U_o(H)$ is weakly dependent on field H [23, 24]. On the other hand, in high field region, the value of $\alpha > 1$, which suggest that flux lines are pinned by collective point defects in this region [24].

AC susceptibility measurement is a technique used to extract superconducting properties like magnetization and critical current etc [25]. Here, the temperature dependence of ac susceptibility of $Nb_2PdS_5$ sample for several magnetic field amplitudes and frequency was performed to understand the AC losses in this sample. Figure 5 (a-b) shows the temperature dependence of real and imaginary part of ac susceptibility for different amplitudes at a fixed frequency of 333Hz. The real part of AC susceptibility shows the transition from superconducting state to normal statei.e. complete penetration of external magnetic flux into the sample [26]. The behaviour of real part in each case clearly shows diamagnetic response below $T$=6.4K, conforming bulk superconductivity. Clearly, the value of $\chi'$ decreases with increasing the amplitude of applied magnetic field. Therefore, in Nb$_2$PdS$_5$ sample, the value of $\chi'$ is dependent on amplitude of magnetic field as well as on temperature. The peak in imaginary part of ac susceptibility occurs at a point, where vortices just reach to the centre of measured sample [27]. Clearly, around the peak temperature $T_p$, the imaginary part shows single sharp peak, which is lower than critical temperature $T_c$. Here, the value of $T_p$ shifts towards lower temperature with increasing amplitude of applied AC magnetic field. Also, the value of full width half maxima of peak enhances with amplitude. Therefore, it can be seen that the value of $T_p$ is strongly dependent upon amplitudes of applied magnetic field.

Figure 5 (c-d) shows temperature dependence of real and imaginary part of ac susceptibility at different ac frequencies. All the measurements were taken at fixed amplitude of 10Oe. Clearly, with varying frequencies, the ac susceptibility curves seem to be independent of the frequency. Apparently, there is no major change in temperature dependence of AC susceptibility curves with various frequencies. Both real and imaginary part of ac susceptibility shows diamagnetic signal at 6K and maximum temperature ($T_p$) at 5.4K. Summarily, the value of ac susceptibility depends on both temperature as well as amplitude of applied magnetic field and is independent of frequency of applied field. The pinning losses are strongly dependent on amplitude of applied AC field, while flux losses depend on frequency of applied magnetic field [28]. In this case, AC susceptibility strongly depends on amplitude and and not on frequency, therefore one can say that in Nb$_2$PdS$_5$ sample mostly the pinning losses occursand the granular coupling is rather strong [28, 29]. To be brave enough one can conjecture that Nb$_2$PdS$_5$ is a well coupled heavily pinned superconductor.

Figure 6 (a) shows the variation of DC magnetization with field (M-H) at several temperatures below superconducting transition. The value of lower critical field ($H_{c1}$) is taken as the field for which M-H curve starts deviating from linear behaviour, as marked in inset of Fig. 6(a). Clearly, the value of lower critical field is decreasing with increase in temperature towards $T_c$. Figure 6 (b) shows the variation of lower critical field ($H_{c1}$) with temperature. Here, $H_{c1}(T)$ curve does not follow general equation given as $H_{c1}(T) = H_{c1}(0)(1-(T/T_c)^2)$ and plotted in Fig. 6(b). This deviation may be the signature of multiband nature of superconductivity [30]. Hence, one can say that the result shown in Fig. 6(b) is in line with the multi band nature of superconductivity in studied $Nb_2PdS_5$ superconductor. Q Zhang et al. claimed multiband nature of superconductivity in $Nb_2PdS_5$ compound as the upper critical field along b axis is linearly dependent on temperature [1]. In our previous report on $Nb_2PdS_5$ compound, the low temperature specific heat measurements revealed multiband superconducting nature with gap values of 1.9 and 6.4, respectively [13]. After this result, the two band superconductivity in $Nb_2PdS_5$ was confirmed via spectroscopy measurement [14]. Interestingly enough, in contrast to ref. 13 and 14, the μSR [7] and penetration depth measurements [31] showed that the $Nb_2PdS_5$ superconductor behaves like weakly coupled BCS single gap superconductor. Present results although support the two band superconductivity as evidenced in ref. 13, 14 but not the ones being from μSR [7] and penetration depth [31]. In short one can say that the superconducting gap symmetry of $Nb_2PdS_5$ superconductor is yet unresolved.

**Conclusion**

Summarily, we have synthesized bulk polycrystalline form of $Nb_2PdS_5$ sample via solid state reaction route. We have studied detailed resistivity and magnetic measurements for as synthesized $Nb_2PdS_5$ sample. The thermally activated flux flow analysis provided the value of activation energy, which decreases with increasing magnetic field. The AC susceptibility is dependent on amplitude, while independent on frequency. This implies prominent pinning losses in $Nb_2PdS_5$ sample. The temperature dependence of lower critical field showed the signature of multiband superconductivity in $Nb_2PdS_5$ sample.

**Acknowledgement**

The authors would like to thank the Director of NPL-CSIR India for his interest in the present work. We acknowledge B. Gahtori for taking SEM images. This work was financially sponsored by UGC –SRF and DAE-SRC projects.

**Figure Captions**

**Figure 1:** Scanning electron microscopy of as synthesized $Nb_2PdS_5$ sample and EDX graph for the selected area of $Nb_2PdS_5$ sample.

**Figure 2:** HRTEM images showing (a) bright field micrograph, (b) layered structure at high magnification, and (c) electron diffraction pattern. Inset in (a) shows an atomic scale image.

**Figure 3:** XPS core level spectra of (a) Niobium, (b) Palladium and (c) Sulphur.

**Figure 4: (a)** Temperature dependent electrical resistivity of $Nb_2PdS_5$ sample under zero magnetic field. Inset of figure **(a)** shows the variation of electrical resistivity of $Nb_2PdS_5$ sample with different magnetic field. **(b)** $\ln\rho(T,H)$ vs $1/T$ for different magnetic field **(c)** Plot of $\ln\rho_0(T,H)$ vs $Uo$(K) **(d)** Thermally activated energy with solid lines fitted using equation $Uo(H) \propto H^{-\alpha}$ for different magnetic fields.

**Figure 5: (a - b)** Temperature dependence real and imaginary part of AC susceptibility at fixed frequency with different amplitudes. **(c - d)** Temperature dependent real and imaginary part of AC susceptibility at fixed amplitude and different frequencies.

**Figure 6: (a)** Magnetic field dependent magnetization of $Nb_2PdS_5$ sample with temperature and Inset shows markedvalue of $H_{c1}$ for $T =2$ K. **(b)** Variation of lower critical field $H_{c1}$ with temperature ($T/T_c$) and solid line represents $H_{c1}$(T) does not follow equation $H_{c1}(0)(1-(T/T_c)^2)$.

**Figures**

Figure 1

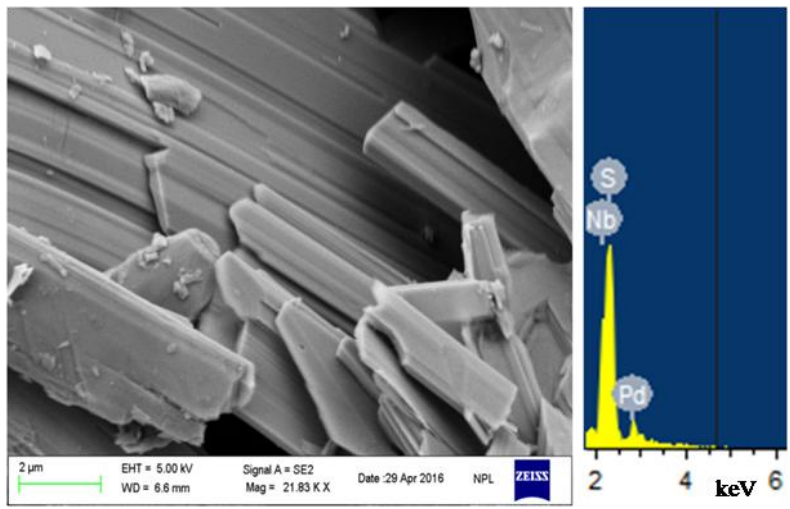

Figure 2

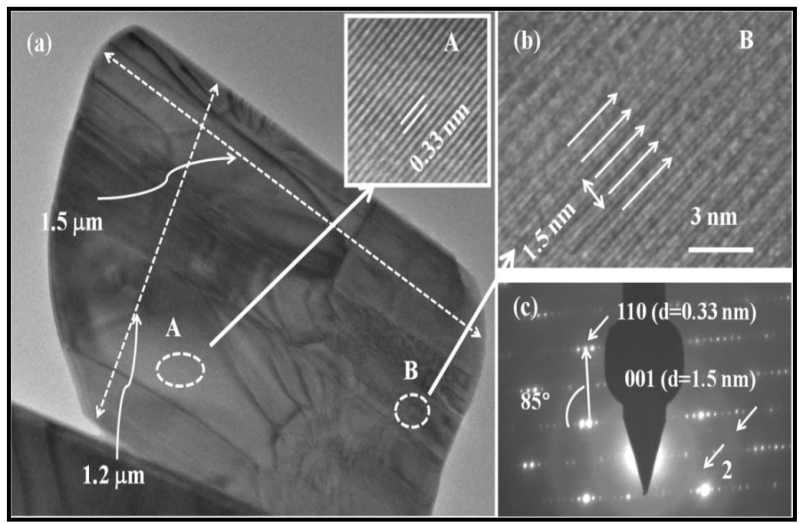

Figure 3

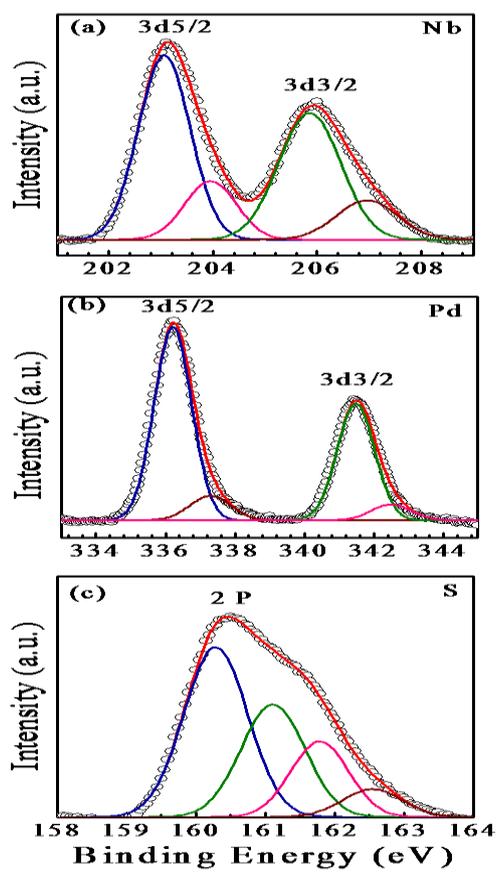

Figure 4

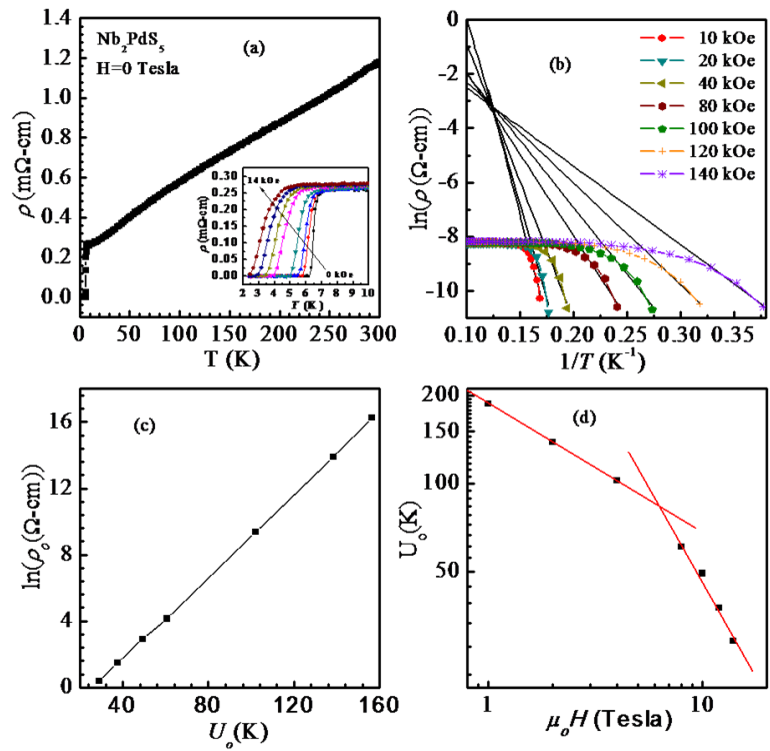

Figure 5

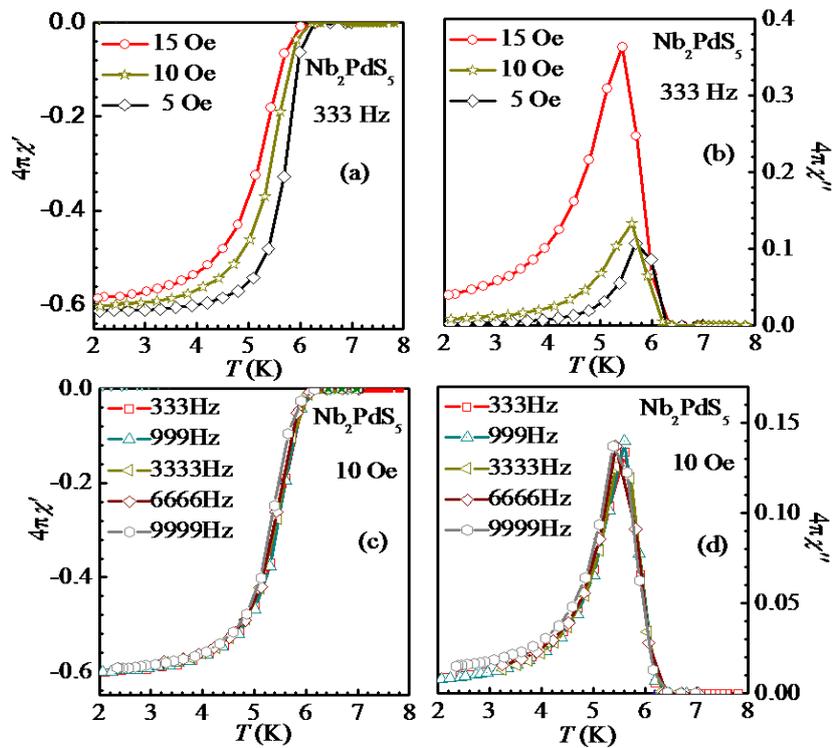

Figure 6

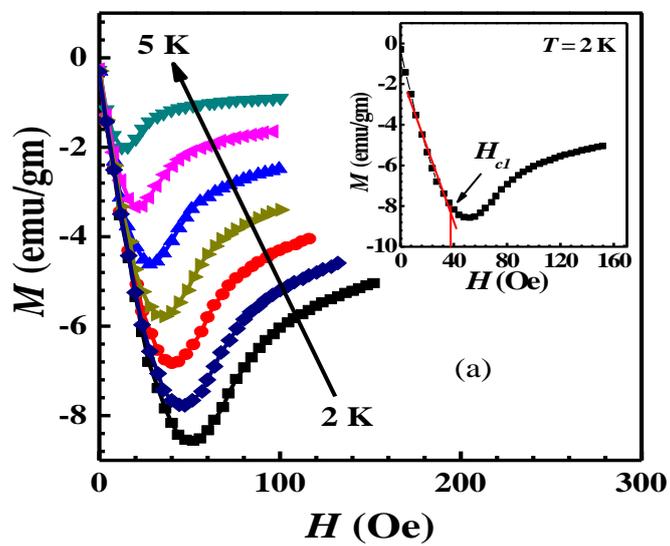

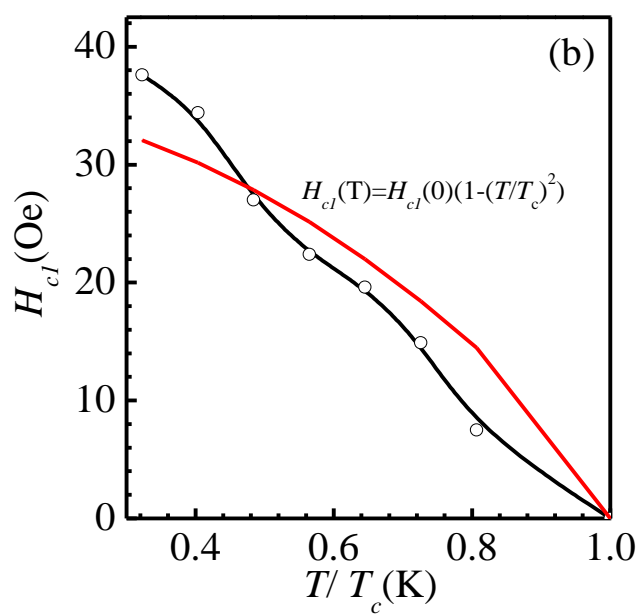